\documentclass[12pt]{article}
\hoffset= - 2cm
\usepackage{amssymb,graphicx}
\usepackage{epstopdf}
\usepackage{amsmath,amsfonts}
\usepackage[normalem]{ulem}
\usepackage{epsfig}
\usepackage[dvipsnames]{xcolor}

\def\eps{\epsilon}

\newcommand{\be}{\begin{equation}}
\newcommand{\ee}{\end{equation}}
\newcommand{\bea}{\begin{eqnarray}}
\newcommand{\eea}{\end{eqnarray}}
\newcommand{\bg}{\begin{gather}}
\newcommand{\eg}{\end{gather}}
\newcommand{\bseq}{\begin{subequations}}
\newcommand{\eseq}{\end{subequations}}

\usepackage{hyperref}
\definecolor{linkcolor}{HTML}{799B03}
\definecolor{urlcolor}{HTML}{799B03}
\hypersetup{pdfstartview=FitH,linkcolor=linkcolor,urlcolor=urlcolor,colorlinks=true}

\makeatletter
\newcommand*{\myfnsymbolsingle}[1]{%
  \ensuremath{%
    \ifcase#1
    \or 
      *%
    \or 
      \dagger
    \or 
      1
    \or 
      2
    \or
      3
    \or
      4
    \or
      5
    \or
      6
    \or
      7
    \or
      8
    \else 
      \@ctrerr
    \fi
  }%
}
\makeatother

\usepackage{alphalph}
\newalphalph{\myfnsymbolmult}[mult]{\myfnsymbolsingle}{}
\renewcommand*{\thefootnote}{%
  \myfnsymbolmult{\value{footnote}}%
}

\begin{document}

\begin{center}
  {\LARGE \bf  Stable non-singular cosmologies in beyond Horndeski theory and disformal transformations}

\vspace{10pt}

\vspace{20pt}
S. Mironov$^{a,b,c,d}$\footnote{sa.mironov\_1@physics.msu.ru},
V. Volkova$^{a}$\footnote{volkova.viktoriya@physics.msu.ru}
\renewcommand*{\thefootnote}{\arabic{footnote}}
\vspace{15pt}

$^a$\textit{Institute for Nuclear Research of the Russian Academy of Sciences,\\
60th October Anniversary Prospect, 7a, 117312 Moscow, Russia}\\
\vspace{5pt}

$^b$\textit{Moscow Institute of Physics and Technology,\\
Institutski pereulok, 9, 141701, Dolgoprudny, Russia}

$^c$\textit{Institute for Theoretical and Mathematical Physics,\\
M.V. Lomonosov Moscow State University, 119991 Moscow, Russia}

$^d$\textit{Kurchatov Institute, ITEP\\
pl. Akademika Kurchatova, 1, 123182 Moscow, Russia}

\end{center}

\vspace{5pt}

\begin{abstract}
In this note we collect, systemise and generalise the existing results
for relations between general Horndeski theories
and beyond Horndeski theories via disformal transformations of metric.
We derive additional disformal transformation rules relating
Lagrangian functions of beyond Horndeski theory and corresponding
Horndeski theory and
demonstrate that some of them become singular at some moments(s) once
one constructs a non-singular cosmological solution in
beyond Horndeski theory that is free from ghost, gradient
instabilities and strong gravity regime during the entire
evolution of the system. The key issue here is that such
solutions are banned in Horndeski theory due to existing
no-go theorem. The proof of singular behaviour of disformal
relations in this case resolves the apparent contradiction
between the fact that Horndeski and beyond Horndeski theories
appear related by field redefinition but describe different
physics in the context of non-singular cosmologies.

\end{abstract}

\section{Introduction}
\label{sec:intro}

Scalar-tensor theories of modified gravity have become
a go-to framework
for addressing various cosmological issues
ranging from initial singularity problem to explaining late time
accelerated expansion.
Today Degenerate Higher order scalar-tensor (DHOST) theories are
the most general class of theories providing a
systemized approach to modifying gravity by adding a scalar
field~\cite{DHOST1,DHOST2,DHOST3}
(see e.g. Ref.\cite{DHOST_LangloisRev} for review).
Despite involving the second derivatives in the Lagrangian, and,
hence, having corresponding equations of motion higher than
second order, DHOST theories are protected from naively
expected Ostrogradsky ghost by a set of degeneracy conditions
imposed on the Lagrangian, which ensure that the number of
propagating degrees of freedom (DOF) is no more than 1 scalar
and 2 tensor modes.

In this note we will focus on two renowned special cases of
DHOST theories, namely, Horndeski
theories~\cite{gal1,gal2,gal3,Horndeski}
and beyond Horndeski or GLPV theories~\cite{Zuma,Gleyzes,Gleyzes2}
(see e.g. Ref.~\cite{KobaRev} for review). The former represent
the most general subclass, where the equations of motion are
manifestly second order in derivatives,
while the latter are historically the first successful generalization of
Horndeski theories, which gave up the restriction of having
second-order equations of motion but still gave rise to
$2+1$ DOFs thanks to degeneracies among the equations.

The equal
number of DOFs in Horndeski theories and their extensions is not
entirely surprising since there is a non-trivial relation
between these theories via the invertible field redefinition,
namely, disformal transformation
of metric~\cite{disformal}
\be
\label{eq:disformal}
\bar{g}_{\mu\nu} = \Omega^2(\pi,X) g_{\mu\nu} +\Gamma(\pi,X) \partial_{\mu}\pi\partial_{\nu}\pi,
\ee
where $\pi$ is a scalar field,
$X=g^{\mu\nu}\partial_{\mu}\pi\partial_{\nu}\pi$, while
$\Omega^2(\pi,X)$ and $\Gamma(\pi,X)$ are arbitrary functions.
Disformal relations between Horndeski and beyond Horndeski subclasses
and phenomenologically viable DHOST theories were discussed in details
e.g. in Refs.~\cite{DHOST2,Gleyzes2} and~\cite{disf0,disf1,disf2}.
In particular, in Ref.~\cite{disf0} it was shown that Horndeski
theories are stable under transformations~\eqref{eq:disformal}
with both $\Omega$ and $\Gamma$ depending on $\pi$ but not $X$. Later
it was shown that allowing $\Gamma$ to be function of $X$ transforms
Horndeski theory into beyond Horndeski class, and, interestingly,
this is exactly how the first example of beyond Horndeski theories was
derived~\cite{Zuma}. Finally, further generalisation with both
$\Omega(\pi,X)$ and $\Gamma(\pi,X)$ allows to generate DHOST
family still starting off with Horndeski theories. Naturally,
as long as no other matter is involved and disformal
transformation~\eqref{eq:disformal} is invertible all these
disformally related theories describe the same physics
\footnote{Recent studies have shows that coupling additional matter
to DHOST theories might be somewhat subtle when preserving
degeneracy is concerned~\cite{Deffayet}.}.

Another exceptional property of DHOST theories and their subclasses
that will be important for us in this note
is their ability to violate the Null Energy Condition (NEC)
%
%
without fatal consequences for the stability of a linearized theory
(see Refs.~\cite{KobaRev,RubakovNEC} for details).
In fact, insofar as gravity is modified NEC is replaced
by the Null Convergence Condition~\cite{TiplerNCC}.
This NEC/NCC-violating feature made Horndeski theories and
their extensions particularly attractive for
non-standard early Universe cosmology, since violating NEC/NCC
is crucial
for constructing cosmological scenarios without initial
singularity like the Universe with a bounce or Universe
starting off with Genesis
(see Ref.~\cite{Khalat} for a mini review).

However, only beyond Horndeski and DHOST theories became
truly successful in pursuit of constructing completely
stable non-singular cosmological
solutions.
''Complete'' stability here means that no
pathological DOFs like ghosts and gradient instabilities are
present in the system during entire evolution followed
from $t\to -\infty$ to
$t\to +\infty$. In particular, unextended Horndeski
theories were ruled out as suitable frameworks for such solutions
by a no-go theorem~\cite{LMR,Koba_nogo}, which states that
any non-singular cosmological solution in Horndeski theory
runs into gradient instabilities at some moment
provided one considers the system from $t\to -\infty$ to
$t\to +\infty$
\footnote{There is a loophole, though: it is possible to
evade the no-go theorem by allowing strong gravity in the past~\cite{Koba_nogo,YuPVA,Nandi}.}.
The no-go breaks down as soon as one goes beyond
Horndeski: this fact was first shown within the effective field
theory approach (EFT) in Refs.~\cite{Cai:2016thi,Creminelli} and
later supported by numerous explicit examples of completely
healthy solutions, see e.g.
Refs.~\cite{bouncegen,CaiBounce,bounceI,genesisGR}.

It might seem contradictory, however,
that while healthy Horndeski and beyond Horndeski theories are
related via a disformal transformation, the former do not
admit completely stable
non-singular cosmologies but the latter do.
Indeed, disformal transformation~\eqref{eq:disformal}
is a field redefinition,
and while it is invertible it cannot change the number of DOFs, hence,
it cannot affect stability of the solution. So the question is how
is it possible that there are non-singular solutions
without instabilities in beyond Horndeski subclass?
This puzzle was resolved in Ref.~\cite{Creminelli}, where
it was found that the disformal transformation, which
turns beyond Horndeski Lagrangian admitting stable solution
into general Horndeski form, becomes singular at some moment of time.
This result was obtained within
the effective field theory (EFT) approach and later confirmed
in the covariant formalism for a quartic subclass of beyond
Horndeski theory~\cite{Khalat}, but not in the most general
case of beyond Horndeski theory including the quintic subclass.

One of the main purposes of this note is to generalize the result of
Ref.~\cite{Khalat} to include both quartic and quintic beyond
Horndeski subclasses, i.e. we show explicitly that
completely stable non-singular cosmological solutions exist only
in those beyond Horndeski theories which are related to Horndeski
subclass by disformal transformations that are singular at some point.
To be precise, singularities appear in the transformation rules
relating the Lagrangian functions, which makes it impossible
to map beyond Horndeski Lagrangian into the Horndeski one in our case.
Another aim of this note is to gather and systemise
the disformal relations between the Lagrangian functions
in Horndeski and beyond Horndeski theories
in a covariant formalism, which were
obtained earlier e.g. in Refs.~\cite{Gleyzes2,disf2}. This is done
in Sec.~\ref{sec:disformal}, where
we also derive some additional relations necessary for our argument.
Then in Sec.~\ref{sec:nogo} we demonstrate that disformal relations
for $X$-derivatives of Lagrangian functions are indeed divergent
if beyond Horndeski theory admits a stable bouncing or Genesis
solution. We conclude in Sec.~\ref{sec:conlusion}.

\section{Disformal relations for Horndeski and GLPV theories}
\label{sec:disformal}

Let us first briefly recall the explicit form of (beyond) Horndeski
Lagrangian and introduce our set of notations:
\be
\label{eq:lagrangian}
\mathcal{L}=\mathcal{L}_2 + \mathcal{L}_3 + \mathcal{L}_4 + \mathcal{L}_5 + {}^{BH}{\cal L}_4 + {}^{BH}{\cal L}_5,
\ee
with
\begin{subequations}
\label{eq:lagrangian_parts}
\begin{align}
&\mathcal{L}_2=F(\pi,X), \quad
\mathcal{L}_3=K(\pi,X)\Box\pi,\\
\label{eq:L4}
&\mathcal{L}_4[G_4]=-G_4(\pi,X)R+2G_{4X}(\pi,X)\left[\left(\Box\pi\right)^2-\pi_{;\mu\nu}\pi^{;\mu\nu}\right], \\
\label{eq:L5}
&\mathcal{L}_5[G_5]=G_5(\pi,X)G^{\mu\nu}\pi_{;\mu\nu}+\frac{1}{3}G_{5X}\left[\left(\Box\pi\right)^3-3\Box\pi\pi_{;\mu\nu}\pi^{;\mu\nu}+2\pi_{;\mu\nu}\pi^{;\mu\rho}\pi_{;\rho}^{\;\;\nu}\right],\\
&{}^{BH}{\cal L}_4[{F}_4]={F}_4(\pi,X)\epsilon^{\mu\nu\rho}_{\quad\;\sigma}\epsilon^{\mu'\nu'\rho'\sigma}\pi_{,\mu}\pi_{,\mu'}\pi_{;\nu\nu'}\pi_{;\rho\rho'}, \\
&{}^{BH}{\cal L}_5[{F}_5]={F}_5(\pi,X)\epsilon^{\mu\nu\rho\sigma}\epsilon^{\mu'\nu'\rho'\sigma'}\pi_{,\mu}\pi_{,\mu'}\pi_{;\nu\nu'}\pi_{;\rho\rho'}\pi_{;\sigma\sigma'},
\end{align}
\end{subequations}
where $R$ and $G_{\mu\nu}$ are Ricci scalar and Einstein
tensor, respectively, while
$\pi$ is a scalar field,
$X=g^{\mu\nu}\pi_{,\mu}\pi_{,\nu}$,
$\pi_{,\mu}=\partial_\mu\pi$,
$\pi_{;\mu\nu}=\nabla_\nu\nabla_\mu\pi$,
$\Box\pi = g^{\mu\nu}\nabla_\nu\nabla_\mu\pi$,
$G_{4X}=\partial G_4/\partial X$, etc.
The terms $\mathcal{L}_2, \mathcal{L}_3, \mathcal{L}_4, \mathcal{L}_5$
in eq.~\eqref{eq:lagrangian} describe general Horndeski theory,
while adding ${}^{BH}{\cal L}_4$ and ${}^{BH}{\cal L}_5$ extend
the theory to beyond Horndeski type.

In what follows we focus on disformal transformation
of the sum of quartic~\eqref{eq:L4} and quintic~\eqref{eq:L5}
subclasses of Horndeski theory
\begin{equation}
\label{eq:starting_lagr}
\mathcal{L}_{H} =  {\cal {L}}_4[\bar{G}_4] + {\cal {L}}_5[\bar{G}_5].
\end{equation}
Here and below we denote metric corresponding to Horndeski theory
as $\bar{g}_{\mu\nu}$, while metric ${g}_{\mu\nu}$ is ascribed
to beyond Horndeski theory.
For our purposes it sufficient to consider
disformal transformation~\eqref{eq:disformal} where $\Omega(\pi,X) = 1$.
Then upon transformation of the metric $\bar{g}_{\mu\nu}$
the Lagrangians
${\cal {L}}_4[\bar{G}_4]$ and ${\cal {L}}_5[\bar{G}_5]$
in eq.~\eqref{eq:starting_lagr} get modified as follows
\begin{equation}
\label{eq:disfL4}
{\cal {L}}_4[\bar{G}_4] =  {\cal L}_2[\hat{F}] + {\cal L}_3[\hat{K}] + {\cal L}_4[\hat{G}_4]  +   {}^{BH}{\cal L}_4[\hat{F}_4] \,,
\end{equation}
\begin{equation}
\label{eq:disfL5}
{\cal {L}}_5[\bar{G}_5] = {\cal L}_2[\tilde{F}] + {\cal L}_3[\tilde{K}] +
{\cal L}_4[\tilde{G}_{4}] + {}^{BH}{\cal L}_4[\tilde{F}_{4}] +
{\cal L}_5[G_5]  +   {}^{BH}{\cal L}_5[F_5] \,,
\end{equation}
where
the beyond Horndeski terms ${}^{BH}{\cal L}_4[\hat{F}_4]$
and ${}^{BH}{\cal L}_5[F_5]$ pop up,
therefore, as expected the resulting Lagrangian belongs
to beyond Horndeski type with metric ${g}_{\mu\nu}$:
\begin{equation}
\label{eq:final_lagr}
\begin{aligned}
&\mathcal{L}_{BH} = \mathcal{L}_2[{F}] + \mathcal{L}_3[{K}]+
\mathcal{L}_4[{G}_4] + \mathcal{L}_5[{G}_5] +
{}^{BH}{\cal L}_4[{F}_{4}] + {}^{BH}{\cal L}_5[{F}_{5}] ,
\end{aligned}
\end{equation}
with the following functions involved
\footnote{Based on the explicit form of $\mathcal{L}_4[G_4]$
in eq.~\eqref{eq:L4} we group similar terms and introduce
the following notation:
$\mathcal{L}_4[\hat{G}_4] + \mathcal{L}_4[\tilde{G}_4] = \mathcal{L}_4[ \hat{G}_4+ \tilde{G}_4] \equiv  \mathcal{L}_4[{G}_4]$.
Similar notations are applied to $\mathcal{L}_2[{F}]$,
$\mathcal{L}_3[{K}]$ and ${}^{BH}\mathcal{L}_4[{F}_4]$. }
\begin{equation}
\label{eq:G4F4hat}
{F} = \hat{F} + \tilde{F}, \qquad {K} = \hat{K} + \tilde{K}, \qquad
{G}_4 = \hat{G}_4 + \tilde{G}_{4}, \qquad
{F}_4 = \hat{F}_{4} + \tilde{F}_{4}.
\end{equation}
Let us note in passing that both eqs.~\eqref{eq:disfL4} and~\eqref{eq:disfL5} show that the disformal transformation applied to
${\cal L}_4$ and ${\cal L}_5$ generates all lower subclasses including ${\cal L}_2$ and ${\cal L}_3$
\footnote{However, since in eq.~\eqref{eq:starting_lagr}
we have knowingly omited ${\cal L}_2$ and ${\cal L}_3$ contributions, we have to do the same in eqs.~\eqref{eq:disfL4} and~\eqref{eq:disfL5}. The latter does not affect the disformal rules for ${G}_4$
and ${G}_5$ that are key for this note}.
This is a common feature of disformal transformations with
$\Gamma(\pi,X)$ applied to Horndeski theories in a covariant
formalism noted already
in Ref.~\cite{Gleyzes2}. Our notations in eq.~\eqref{eq:G4F4hat}
aim to emphasise that both $\mathcal{L}_4$ and ${}^{BH}\mathcal{L}_4$
in~\eqref{eq:final_lagr} have two distinct contributions
with different disformal transformations rules for corresponding
Lagrangian functions $\hat{G}_4$, $\hat{F}_4$ and $\tilde{G}_4$, $\tilde{F}_4$
as we show below.

Now let us revisit the existing results for disformal relations
between the Lagrangian functions and derive the missing ones.
The relations between the original functions $\bar{G}_4$
and $\bar{G}_5$ in Horndeski theory and the new $\hat{G}_4$
and $G_5$ in beyond Horndeski theory in the covariant
approach were already found in Refs.~\cite{Gleyzes2,disf2}
and read as follows:
\begin{eqnarray}
\label{eq:newG4G5}
\bar{G}_4 (\pi, \bar{X}) &=& \frac{\hat{G}_4 (\pi, {X})}  {\sqrt{1 + X \Gamma}}\,,
\qquad
\bar{G}_5(\pi, \bar{X}) = \int G_{5X} (\pi, {X}) \sqrt{1 + \Gamma X} dX\,,
\qquad
\bar{X} = \frac{X}{1 + X \Gamma} \,.
\end{eqnarray}
After disformal transformation the following combinations
of $\hat{G}_4$, $G_5$ and $\Gamma$ in the
transformed Lagrangians~\eqref{eq:disfL4}--\eqref{eq:disfL5}
comprise
the beyond Horndeski functions $\hat{F}_4$ and $F_5$:
\begin{subequations}
\label{eq:newF4F5}
\begin{align}
\label{eq:newF4}
&\hat{F}_4 = \dfrac{\Gamma_X \left( \hat{G}_4-2 X \hat{G}_{4X}\right)}{1-\Gamma_X X^2},
\\
\label{eq:newF5}
&F_5 =-\dfrac{\Gamma_X\; G_{5X} X }{3(1-\Gamma_X X^2)}.
\end{align}
\end{subequations}
At this point modulo ${F}$ and ${K}$ the only new functions
left in eq.~\eqref{eq:G4F4hat} to be related with the original ones
are $\tilde{G}_4$ and $\tilde{F}_4$.
As we mentioned above disformal transformation of ${\cal L}_5[\bar{G}_5]$
generally
involves all lower subclasses, and according to eq.~\eqref{eq:disfL5}
$\tilde{G}_4$ and $\tilde{F}_4$ belong to such generated $\mathcal{L}_4$
and ${}^{BH}\mathcal{L}_4$, respectively.
Hence, both $\tilde{G}_4$ and $\tilde{F}_4$ are given
by some combinations in terms of $\Gamma$ and the new $G_5$:
\begin{subequations}
\label{eq:tildeG4F4}
\begin{align}
\label{eq:tildeG4}
&\tilde{G}_4 (\pi,X) = \dfrac{X}{4} \int \left[ \int G_{5\pi X} X^{1/2} dX\right] X^{-3/2} dX - \Bar{\Bar{G}}_4(\pi,X)
\;, \\
\label{eq:tildeF4}
&\tilde{F_4}(\pi,X) = - \dfrac{\Gamma_X \left( \Bar{\Bar{G}}_4-2 X \Bar{\Bar{G}}_{4X}\right)}{1-\Gamma_X X^2} \;,
\end{align}
\end{subequations}
where
\begin{equation*}
\Bar{\Bar{G}}_4(\pi,X) = \dfrac{X}{4 \sqrt{1 +\Gamma X}}
\int \left[ \int G_{5\pi X} X^{1/2} dX\right]  \dfrac{1 - \Gamma_X X^2}{\sqrt{1 +\Gamma X}} X^{-3/2} dX \;,
\end{equation*}
was introduced to highlight the structure of the
transformation rules~\eqref{eq:tildeG4F4}. Eq.~\eqref{eq:tildeG4} together with $\hat{G}_4$ in eq.~\eqref{eq:newG4G5} define the relation between the original $\bar{G}_4$ and the new $G_4 = \hat{G}_4 + \tilde{G}_4$.
Then the explicit form of ${F}_4 = \hat{F}_4 + \tilde{F}_4$
in eq.~\eqref{eq:final_lagr} follows from eqs.~\eqref{eq:newF4}
and~\eqref{eq:tildeF4} with integrating by parts where it is necessary:
\begin{equation}
\label{eq:hatF4}
{F}_4 = \dfrac{\Gamma_X}{1-\Gamma_X X^2}\left( {G}_4
-2 X {G}_{4X} + \dfrac12 G_{5\pi}X \right).
\end{equation}
So in result
eqs.~\eqref{eq:newG4G5} together with
explicitly derived in this note
eqs.~\eqref{eq:newF5},~\eqref{eq:tildeG4}
and~\eqref{eq:hatF4} completely define the
disformal relation between Lagrangian functions
in $\mathcal{L}_H$~\eqref{eq:starting_lagr}
and $\mathcal{L}_{BH}$~\eqref{eq:final_lagr}.
Let us note that it follows immediately
from eqs.~\eqref{eq:newF5} and~\eqref{eq:hatF4} that both
${F}_4$ and $F_5$ vanish for $\Gamma=\Gamma(\pi)$ in
full agreement with the fact that Horndeski theories are
stable under disformal transformations with $\Gamma_X = 0$~\cite{disf0}.

Another important by-product of revisiting the disformal relations above
is the ability to explicitly derive a specific constraint on
${F}_4$ and $F_5$ in the resulting beyond Horndeski theory
\footnote{The aforementioned constraint has been discussed earlier
in literature and given without derivation e.g.
in Ref.~\cite{F4F5relation}. }:
\begin{equation}
\label{eq:F4F5relation}
{F}_4 \; G_{5X} X = -3 F_5 \;\left[{G}_4 - 2 X {G}_{4X} + \frac12 G_{5\pi} X\right],
\end{equation}
which immediately
follows from eqs.~\eqref{eq:newF5} and~\eqref{eq:hatF4}.
In particular, the relation~\eqref{eq:F4F5relation}
is in line with the fact that beyond Horndeski with arbitrary
${F}_4$ and $F_5$ cannot be disformally transformed
into Horndeski theory~\cite{Gleyzes2,disf1,disf2}.

Finally, we derive the relations between the original functions $\bar{G}_{4\bar{X}}$
and $\bar{G}_{5\bar{X}}$ and the new ones, which immediately follow from
eqs.~\eqref{eq:newG4G5}:
\begin{eqnarray}
\label{eq:G4XG5X}
&\bar{G}_{4\bar{X}} = \dfrac{\partial \bar{G}_4}{\partial \bar{X}}  = \left( \hat{G}_4 (1+\Gamma X)- \dfrac{1}{2} \hat{G}_{4}(\Gamma + X \Gamma_X) \right)\dfrac{\sqrt{1+\Gamma X}}{1-\Gamma_X X^2}, \\
\nonumber
&\bar{G}_{5\bar{X}} = \dfrac{\partial \bar{G}_5}{\partial \bar{X}} =
G_{5X} \dfrac{(1+\Gamma X)^{5/2}}{1 - \Gamma_X X^2}.
\end{eqnarray}
In the next section based on these relations we will demonstrate
that both
$\bar{G}_{4\bar{X}}$ and $\bar{G}_{5\bar{X}}$ which enter the Lagrangian
$\mathcal{L}_H$~\eqref{eq:starting_lagr}
inevitably become singular at some moment provided the corresponding
beyond Horndeski theory~\eqref{eq:final_lagr} admits
stable non-singular cosmological solutions.

\section{No-go theorem and disformal transformations}
\label{sec:nogo}

Let us start with a quick revision of a linearized theory for
Horndeski theories and beyond Horndeski theories~\eqref{eq:lagrangian}.
The corresponding quadratic action for perturbations about
a cosmological FLRW background has the same structure for both theories.
In the unitary gauge where the propagating DOFs are curvature
perturbation $\zeta$ and two tensor modes $h^T_{ij}$, the action reads
(see e.g. Ref.~\cite{KobaRev} for details):
\begin{equation}
\label{eq:unconstrained_action}
S=\int\mathrm{d}t\mathrm{d}^3x \;a^3
\left[\dfrac{\mathcal{{G}_T}}{8}\left(\dot{h}^T_{ij}\right)^2
-\dfrac{\mathcal{F_T}}{8a^2}\left(\partial_k h_{ij}^T\right)^2
+\mathcal{G_S}\dot{\zeta}^2
-\mathcal{F_S}\dfrac{(\partial_k\zeta)^2}{a^2}\right],\\
\end{equation}
with an overdot standing for the derivative w.r.t. time $t$,
scale factor $a$
and the following coefficients involved
\bea
\label{eq:GS_setup}
&&\mathcal{G_S}=\dfrac{\Sigma\mathcal{{G}_T}^2}{\Theta^2}+3\mathcal{{G}_T},\\
\label{eq:FS_setup}
&&\mathcal{F_S}=\dfrac{1}{a}\dfrac{\mathrm{d}}{\mathrm{d}t}
\left[ \dfrac{a\left(\mathcal{{G}_T}+\mathcal{D}\dot{\pi}\right)\mathcal{{G}_T}}{\Theta}\right]
-\mathcal{F_T},
\eea
where $\cal{G_T}$, $\cal{F_T}$, $\cal{D}$, $\Theta$ and
$\Sigma$ are given in terms of Lagrangian functions
\begin{subequations}
\label{eq:coeff_setup}
\begin{align}
\label{eq:GT_coeff_setup}
&\mathcal{G_T}=2G_4-4G_{4X}X+G_{5\pi}X-2HG_{5X}X\dot{\pi} + 2F_4X^2+6HF_5X^2\dot{\pi},
\\
&\mathcal{F_T}=2G_4-2G_{5X}X\ddot{\pi}-G_{5\pi}X,\\
\label{eq:D_coeff_setup}
&\mathcal{D}=-2F_4X\dot{\pi}-6HF_5X^2,\\
&\Theta=-K_XX\dot{\pi}+2G_4H-8HG_{4X}X-8HG_{4XX}X^2+G_{4\pi}\dot{\pi}+2G_{4\pi X}X\dot{\pi}-5H^2G_{5X}X\dot{\pi}\nonumber\\
&-2H^2G_{5XX}X^2\dot{\pi}+3HG_{5\pi}X+2HG_{5\pi X}X^2
+10HF_4X^2+4HF_{4X}X^3+21H^2F_5X^2\dot{\pi}
 \nonumber\\
 &
 +6H^2F_{5X}X^3\dot{\pi},
 \label{eq:Theta_coeff_setup}
\\
&\Sigma=F_XX+2F_{XX}X^2+12HK_XX\dot{\pi}+6HK_{XX}X^2\dot{\pi}-K_{\pi}X-K_{\pi X}X^2-6H^2G_4
\nonumber\\
&+42H^2G_{4X}X+96H^2G_{4XX}X^2+24H^2G_{4XXX}X^3-6HG_{4\pi}\dot{\pi}-30HG_{4\pi X}X\dot{\pi}
\nonumber\\
 \nonumber
&-12HG_{4\pi XX}X^2\dot{\pi}+30H^3G_{5X}X\dot{\pi}+26H^3G_{5XX}X^2\dot{\pi}+4H^3G_{5XXX}X^3\dot{\pi}-18H^2G_{5\pi}X\\
 \nonumber
&-27H^2G_{5\pi X}X^2-6H^2G_{5\pi XX}X^3-90H^2F_4X^2-78H^2F_{4X}X^3-12H^2F_{4XX}X^4\\
&-168H^3F_5X^2\dot{\pi}-102H^3F_{5X}X^3\dot{\pi}-12H^3F_{5XX}X^4\dot{\pi}.
\label{eq:Sigma_coeff_setup}
\end{align}
\end{subequations}
The case of Horndeski theory
is recovered immediately once $F_4=F_5 = 0$ in
eqs.~\eqref{eq:coeff_setup}.

To have a cosmological solution that is free
from ghost and gradient instabilities one
has to satisfy the following inequalities:
\begin{equation}
\label{eq:stability}
\mathcal{G_T}, \mathcal{F_T} > \epsilon > 0, \qquad
\mathcal{G_S},\mathcal{F_S} > \epsilon > 0,
\end{equation}
where $\epsilon$ is a positive constant which ensures that there is no
naive strong coupling, i.e. $\mathcal{{G}_{S,T}} \not\to 0$
and/or $\mathcal{{F}_{S,T}} \not\to 0$.
This is a strong version of
stability constraints adopted here to avoid considering the case of
$\mathcal{{G}_{S,T}} \to 0$
and/or $\mathcal{{F}_{S,T}} \to 0$ which might help evade the no-go
without encountering the strongly coupled regime, see Refs.~\cite{Koba_nogo,YuPVA} for details.

The no-go theorem~\cite{LMR,Koba_nogo} in Horndeski theory
is based on the gradient stability constraint~\eqref{eq:FS_setup}
in cosmologies with a scale factor $a \neq 0$:
\begin{equation}
\label{eq:nogo}
\dfrac{\mathrm{d}}{\mathrm{d}t}\left[\dfrac{a\mathcal{G_T}(\mathcal{G_T}+\cal{D}\dot{\pi})}{\Theta}\right] =
a \left(\mathcal{F_S} +\mathcal{F_T}\right) > \eps > 0,
\end{equation}
where $\mathcal{D}=0$ for Horndeski theory,
see eq.~\eqref{eq:D_coeff_setup}. To have complete stability
one requires that the constraint~\eqref{eq:nogo} holds at all times.
Then according to eq.~\eqref{eq:nogo}, the coefficient on the
left-hand side
\begin{equation}
\label{eq:xi_H}
\xi=\dfrac{a\mathcal{G_T}^2}{\Theta}
\end{equation}
has to be a monotonously growing function with slope bounded
from below by $\epsilon$, which means that
$\xi \to -\infty$ as $t \to -\infty$ and $\xi \to +\infty$
as $t \to +\infty$, i.e. $\xi$ has to cross zero at some moment(s)
of time.
However, for $a > 0$ and $\mathcal{G_T} >\epsilon>0$
(due to no ghost constraint in the tensor sector, see
eq.~\eqref{eq:stability}) the only option for $\xi$ to cross zero
is when $\Theta \to \infty$, which corresponds to a singularity
in the classical solution.
Thus, non-singular cosmological solutions in Horndeski theory
cannot satisfy the stability conditions~\eqref{eq:stability}
at all times.

The situation changes as soon as $\mathcal{D} \neq 0$ in
eq.~\eqref{eq:nogo}, which is the case for beyond Horndeski theory:
while $\mathcal{G_T}$ still has to be positive due to the
no-ghost condition~\eqref{eq:stability}, the combination
$(\mathcal{G_T} + \cal{D})$ is unconstrained and may take any
values including zero, so that
\begin{equation}
\label{eq:tildeXi}
\tilde\xi = \dfrac{a\mathcal{G_T}(\mathcal{G_T}+\cal{D}\dot{\pi})}{\Theta}
\end{equation}
on the left-hand side in eq.~\eqref{eq:nogo}
can safely cross zero and grow monotonously during entire evolution.
Therefore, in beyond Horndeski theories it is possible to construct
a completely stable non-singular cosmological solution.
Let us now demonstrate
that in fact right at the moment when
$\tilde\xi \sim (\mathcal{G_T} + \mathcal{D}\dot{\pi})$ crosses zero
in beyond Horndeski theory, the original functions
$\bar{G}_{4\bar{X}}$ and $\bar{G}_{5\bar{X}}$ in a
disformally related Horndeski Lagrangian
$\mathcal{L}_H$~\eqref{eq:starting_lagr}
hit singularity.

First, we express $\Gamma_X$ using a linear combination of
eqs.~\eqref{eq:newF5} and~\eqref{eq:hatF4}
\footnote{Here the linear combination is explicitly the sum of
eqs.~\eqref{eq:newF5} and~\eqref{eq:hatF4} multiplied by
$2 H X^2 \dot{\pi}$ and $2 X^2$, respectively.}
and cast it in terms of coefficients~\eqref{eq:coeff_setup}
from the quadratic action:
\begin{equation}
\label{eq:gamma_x}
\Gamma_X = -\dfrac{\cal{D}\dot{\pi}}{X^2 \mathcal{G_T}}.
\end{equation}
Both transformation rules~\eqref{eq:G4XG5X} for $\bar{G}_{4\bar{X}}$
and $\bar{G}_{5\bar{X}}$ involve the same denominator,
which in terms of $\mathcal{D}$ and $\mathcal{G_T}$ reads:
\begin{equation}
\label{eq:gamma_x_frac}
\dfrac{1}{1- \Gamma_X X^2}= \dfrac{\mathcal{G_T}}{\mathcal{G_T} + \cal{D}\dot{\pi}}\;.
\end{equation}
It follows from eq.~\eqref{eq:gamma_x_frac} that the denominator of
the transformation rules~\eqref{eq:G4XG5X} goes through zero at the
same moment as $\tilde\xi$ in eq.~\eqref{eq:tildeXi} does.
This means that both Lagrangian
functions $\bar{G}_{4\bar{X}}$ and $\bar{G}_{5\bar{X}}$
diverge at that moment of time.

To sum up, once one goes beyond Horndeski to evade the no-go
theorem by having $\mathcal{D}\neq 0$, it is possible to comply
with the requirement for $\tilde\xi$ in eq.~\eqref{eq:nogo}.
The latter implies vanishing $(\mathcal{G_T} + \cal{D}\dot{\pi})$
at some moment(s) of time. And, thus, in this case beyond Horndeski
and Horndeski theories are related by singular transformation
rules~\eqref{eq:G4XG5X}. So in fact there is no contradiction
between the no-go theorem and existence of completely stable
cosmologies in seemingly disformally related theories.

\section{Conclusion}
\label{sec:conlusion}

Even though disformal transformations within scalar-tensor theories of
modified gravity are a highly-developed topic, in this note we aimed
to collect, somewhat systemise and generalize the existing results
in the covariant formalism for beyond Horndeski theories.
In this way, we have addressed a specific issue of disformal relation
between Horndeski and beyond Horndeski theories in the context
of constructing
healthy non-singular cosmological solutions like bouncing
Universe or the Universe with Genesis.
In particular, disformal relation between the theories implies
similar physics behind both of them, but at the same time there
is the no-go theorem which states that completely stable
non-singular cosmologies exist only in beyond Horndeski theory
but not in Horndeski subclass.
The resolution of this apparent contradiction is similar
to that suggested in Ref.~\cite{Khalat} for
the quartic subclass of beyond Horndeski theories: the price of
evading the no-go theorem is singularity in disformal relations
between the Lagrangian functions of corresponding Horndeski and
beyond Horndeski theories.
In this note we have proved the
latter statement for the case of quintic beyond Horndeski
subclass, which generalizes the existing result and completes
the argument.

\section{Acknowledgements}
Authors are grateful to V. Rubakov for very useful discussions.
The work on Sec.~\ref{sec:disformal} of this paper has been supported
by Russian Science Foundation grant 19-12-00393, while the
part of the work in Sec.~\ref{sec:nogo} has been
supported by the Foundation for the Advancement of
Theoretical Physics and Mathematics “BASIS”.


\end{document}